\documentclass[conference]{IEEEtran}
\IEEEoverridecommandlockouts
\usepackage{cite}
\usepackage{amsmath,amssymb,amsfonts}
\usepackage{graphicx}
\usepackage{textcomp}
\usepackage{xcolor}
\usepackage{url}
\usepackage{multicol}
\usepackage{colortbl}
\usepackage[normalem]{ulem}
\useunder{\uline}{\ul}{}

\usepackage{algpseudocode}

\usepackage[normalem]{ulem}
\useunder{\uline}{\ul}{}
\usepackage{graphicx}
\usepackage{subcaption}
\usepackage{mwe}
\usepackage{comment}
\usepackage{amsmath}
\usepackage{algpseudocode}
\usepackage{algorithm}
\usepackage{lipsum}
\usepackage[export]{adjustbox}
\usepackage{amsthm}

\algnewcommand\algorithmicforeach{\textbf{for each}}
\algdef{S}[FOR]{ForEach}[1]{\algorithmicforeach\ #1\ \algorithmicdo}

\def\by{{\mathbf y}}
\def\bI{{\mathbf I}}
\def\bPhi{{\mathbf \Phi}}
\def\bPsi{{\mathbf \Psi}}

\def\balpha{{\mathbf \alpha}}

\def\br{{\mathbf r}}
\def\bn{{\mathbf n}}
\def\bk{{\mathbf k}}
\def\boldeta{{\pmb \eta}}
\usepackage{amsmath,amssymb,amsfonts}
\def\CC{\mathbb{C}}
\def\balpha{\boldsymbol{\alpha}}
\def\Phib{\tilde{\bPhi}}
\def\zerovec{{\pmb{0}}}
\def\bvarepsilon{{\mathbf \varepsilon}}

\def\cCN{\mathcal{CN}}

\def\bQ{{\mathbf Q}}
\def\sigmaalpha{\sigma_\alpha^2}

\def\upredmean{{\Tilde{u}}}

\def\ypredvar{{\tau^2}}

\def\Sigmaalpha{{\boldsymbol{\Sigma}_\alpha}}
\def\Sigmaalphainv{{\boldsymbol{\Sigma}_\alpha^{-1}}}
\def\uhat{\hat{u}}
\def\tu{\Tilde{u}}
\def\cB{\mathcal{B}}

\def\minwrt[#1]{\underset{#1}{\mathrm{minimize }}}
\def\argminwrt[#1]{\underset{#1}{\mathrm{arg\;\min }}}

\floatstyle{ruled}
\newfloat{algorithm}{t}{lop}
\floatname{algorithm}{Algorithm}

\newcommand{\bu}{\mathbf{u}}

\newcommand{\bC}{\mathbf{C}}

\newcommand{\bxi}{\boldsymbol{\xi}}

\newcommand{\RR}{\mathbb{R}}

\def\BibTeX{{\rm B\kern-.05em{\sc i\kern-.025em b}\kern-.08em
    T\kern-.1667em\lower.7ex\hbox{E}\kern-.125emX}}
\begin{document}

\title{Boundary-Informed Sound Field Reconstruction}

\author{\IEEEauthorblockN{David Sundström\IEEEauthorrefmark{2}, Filip Elvander\IEEEauthorrefmark{1}, and Andreas Jakobsson\IEEEauthorrefmark{2}
}
\IEEEauthorblockA{\IEEEauthorrefmark{1}Dept. of Information and Communications Engineering, Aalto University, Finland\\
}
\IEEEauthorblockA{\IEEEauthorrefmark{2}Dept. of Mathematical Sciences, Lund University, Sweden
}
}

\maketitle

\begin{abstract}
We consider the problem of reconstructing the sound field in a room using prior information of the boundary geometry, represented as a point cloud. 
In general, when no boundary information is available, an accurate sound field reconstruction over a large spatial region and at high frequencies requires numerous microphone measurements. On the other hand, if all geometrical and acoustical aspects of the boundaries are known, the sound field could, in theory, be simulated without any measurements. In this work, we address the intermediate case, where only partial or uncertain boundary information is available.
This setting is similar to one studied in virtual reality applications, where the goal is to create a perceptually convincing audio experience. In this work, we focus on spatial sound control applications, which in contrast require an accurate sound field reconstruction. Therefore, we formulate the problem within a linear Bayesian framework, incorporating a boundary-informed prior derived from impedance boundary conditions. The formulation allows for joint optimization of the unknown hyperparameters, including the noise and signal variances and the impedance boundary conditions. Using numerical experiments, we show that incorporating the boundary-informed prior significantly enhances the reconstruction, notably 
even when only a few hundreds of boundary points are available or when the boundary positions are calibrated with an uncertainty up to $1$ dm. 

\end{abstract}


\vspace{3mm} 

\section{Introduction}
\label{sect:introduction}

Applications involving sound field control in a defined region, such as spatial active noise control \cite{Koyama2021} and sound zone generation \cite{wu2010spatial}, depend on accurately reconstructing the sound field from nearby microphone measurements.
In general, this is a challenging problem, in particular 
when a large region and high frequencies are of interest. 
In such settings, the number of parameters required to accurately represent the sound field is typically 
orders of magnitude greater than the number of available microphone measurements, necessitating the use of some form of regularization to allow for a unique solution.

Given the importance of the problem, various ways to introduce regularizing prior information has been studied in the recent literature \cite{caviedes2021gaussian}.
For example, when using a linear setting with plane wave basis functions \cite{horiuchi2021kernel}, free-field Greens functions \cite{antonello2017room}, and/or spatial Fourier basis functions \cite{williams1999fourier}, the most common approach, by far, is using Tikhonov regularization \cite{tikhonov1977solutions}, corresponding to an assumption of i.i.d. normal priors on the coefficients. While these assumptions are well suited for a diffuse sound field \cite{ueno2017sound}, sparse priors has also been used to promote a directional sound field \cite{verburg2018reconstruction,horiuchi2021kernel}. 
Additionally, for nonlinear models or linear models with basis functions that  by construction do not satisfy the Helmholtz or wave equation, regularizers have been introduced to enforce pointwise adherence to the differential equation \cite{damiano2024zero,sundstrom2024sound}. Regardless of the type of regularization, the information provided by the microphone measurements is limited. In contrast, in an idealized scenario where the source position, source signal, speed of sound, boundary geometry, and acoustic properties are all known, the sound field can be computed as a forward problem, which is, for instance, exploited in simulation-based reconstruction methods \cite{veronesi1989digital}. 

Such methods have been extended to both time- and frequency domain formulations for the settings where the boundary conditions are either unknown \cite{nava2009situ,antonello2014identification}, partially unknown \cite{schmid2023bayesian}, the source positions are unknown \cite{ahn2023novel}, or for when both the source position and boundary conditions are unknown \cite{bertin2016joint}. Regrettably, such simulation-based formulations require complete and precise calibration of the boundary geometry for the forward-problem to be meaningful. Even if possible to measure, it would typically be a both expensive and time-consuming process to obtain this information. 

In contrast, partial information of the boundary-surfaces can easily be obtained using consumer-grade devices, such as a smartphone camera capturing just a few photos of a room \cite{leroy2024grounding}. Although the reconstructed point cloud may be coarse, uncertain, or incomplete,  it can still provide valuable information that may be exploited for sound field reconstruction. 
The use of such partial boundary data has recently been explored in virtual reality applications \cite{chen2023everywhere,liang2024av,liang2023neural,majumder2022few}, demonstrating successful reconstruction of perceptually convincing binaural signals when evaluated using metrics such as reverberation time, speech intelligibility, and early decay time.
However, in sound field control applications, as studied here, precise reconstruction of the sound field is essential.

Aiming to improve the accuracy of the sound field reconstruction, this work proposes to regularize the sound field reconstruction problem using a three-dimensional (3D) point cloud that partially represents the boundaries of the room.
To do so, we formulate the problem in a linear Bayesian framework, deriving a prior distribution for the linear coefficients from impedance boundary conditions \cite{kuttruff2016room}. Unknown hyperparameters, such as the signal and noise variances and the boundary-related parameters, are jointly optimized. When the hyperparameters are known, the method has similar computational complexity as other Gaussian process-based estimators \cite{caviedes2021gaussian}, and can thus be used in downstream applications related to sound field control (see for example \cite{Koyama2021,brunnstrom2022variable}).

\newpage 

\section{Problem formulation}
\label{sect:signal_model}
Consider a sound field $u: \RR^3\rightarrow \CC$, for a given frequency $\omega$, that is measured using $M$ microphones, such that 
\begin{equation}
    \by = \bu + \varepsilon,
    \label{eq:measurement_model}
\end{equation}
where $\bu = [u(\br_1), \dots, u(\br_M)]$ denotes the sound field at the microphone positions $\br_m\in \Omega \subseteq\RR^3$, for $m = 1,\dots, M$, and $\varepsilon = [\varepsilon_1, \dots, \varepsilon_M]$ is the measurement noise, here assumed to be well modelled as an independent circularly symmetric zero-mean Gaussian noise, i.e., 
\begin{equation}
    \bvarepsilon_i \sim \cCN \left(\zerovec, \sigma^2\bI \right),
\end{equation}
where $\bI$ denotes the identity matrix of appropriate dimension. The region $\Omega$ is assumed to be source-free, such that the sound field satisfies the homogeneous Helmholtz equation
\begin{align}
    \nabla^2u(\br)+k^2u(\br) = 0  \quad &\forall \quad \br\in \Omega, \label{eq:homogeneous_helmholtz}
\end{align}
where $k = {\omega}/{c}$ denotes the wave number and $c$ the speed of sound.
The sound field is here represented using a superposition of plane wave functions, also known as the {Herglotz wave functions} \cite{colton1998inverse}, defined as

\begin{equation}
    u(\br) = \int_{\boldeta \in \mathcal{S}_2} \tu(\boldeta)e^{ik\boldeta^T\br}d\boldeta,
    \label{eq:Herglotz}
\end{equation}
where $\tu: \mathcal{S}_2 \rightarrow \CC$ is the complex source distribution on the unit sphere $\mathcal{S}_2$ in $\RR^3$. The representation in \eqref{eq:Herglotz} satisfies \eqref{eq:homogeneous_helmholtz} and any solution to \eqref{eq:homogeneous_helmholtz} can be arbitrarily well approximated by the Herglotz wave functions \cite{ueno2021directionally}. 

Furthermore, we assume a subset of the boundary of the region, denoted $\cB \subseteq\delta\Omega$, to consist of a locally reacting surface with acoustical properties independent of the angle of incidence. Under these assumption, the acoustical properties of the boundaries can be characterized in terms of the specific impedance $\beta: \cB \rightarrow \CC$ using the impedance boundary conditions
\cite{kuttruff2016room}
\begin{align}
    \beta(\br)\bn(\br) \cdot \nabla u(\br) + i k u(\br) = 0\quad &\forall \quad \br \in \cB \label{eq:robin_bc},
\end{align}
where $\bn: \RR^3 \rightarrow \mathcal{S}_2$ is the outward normal to the surface. In this work, we consider the problem of reconstructing the sound field in the region $\Omega$, given the measurements in \eqref{eq:measurement_model}, the microphone positions $\{\br_m\}_{m = 1,\dots, M}$, the wave number $k$, and  the set of boundary points $\cB$ with corresponding normals.

\section{Proposed method}
\label{sect:method}
In order to introduce the partial boundary information in the sound field reconstruction problem, we here employ a Bayesian problem formulation \cite{rasmussen2005gaussian, caviedes2021gaussian}. 
Let 
\begin{equation}
    \by = \bPhi \balpha + \varepsilon,
\end{equation}
be the discretized Herglotz representation of the measurement model in \eqref{eq:measurement_model}, where $\balpha = [\balpha_1, \dots, \balpha_P]^T \in \CC^P$ denotes the unknown vector of coefficients and $[\bPhi]_{m,p} = \phi_p(\br_m)$, where $\phi_p(\br) = e^{i\bk_p \br}$ is a plane wave with $\bk_p = k\boldeta_p$, with $\boldeta_p \in \mathcal{S}_2$ denoting the direction of arrival. The likelihood of observing $\by$, given the parameters, is given by
$
    \by| \sigma^2, \balpha \sim \cCN \left(\bPhi\alpha, \sigma^2\bI \right).
$
A naive approach to estimate the sound field would be to maximize the likelihood function of $\balpha$, which corresponds to solving the problem
\begin{equation}
    \minwrt[\balpha \in \CC] \quad \frac{1}{2 \sigma^2}||\by - \bPhi \balpha ||_2^2.
    \label{eq:least_squares}
\end{equation}
However, this problem is typically ill-posed since $M\ll P$ in general. Instead, prior information about the sound field can be included by maximizing the {\em a posteriori} distribution, given by $p(\balpha | \by,\sigma^2)\propto p(\by|\balpha,\sigma^2)p(\balpha)$. When the prior distribtuion is a zero-mean circularly symmetric complex normal distribution, i.e., 
\begin{equation}
    \balpha \sim \cCN(\zerovec, \Sigmaalpha),
    \label{eq:alpha_prior}
\end{equation}
the maximum {\em a posteriori} estimate is given as the solution to 
\begin{equation}
    \balpha_{\mathrm{MAP}}= \argminwrt[\balpha \in \CC^P] \quad \frac{1}{2 \sigma^2}||\by - \bPhi \balpha ||_2^2 + \balpha^H\Sigmaalphainv\balpha,
\end{equation}
which has the closed form expression
\begin{equation}
    \balpha_{\mathrm{MAP}} = \frac{1}{\sigma^2}\left(\frac{1}{\sigma^2}\bPhi^H\bPhi + \Sigmaalphainv \right)^{-1}\bPhi^H \by.
    \label{eq:alpha_MAP}
\end{equation}
Using \eqref{eq:alpha_MAP}, the predictive distribution of the sound field at position $\br$
is given by $\uhat(\br) \sim \cCN\left(\upredmean(\br),\ypredvar(\br) \right)$, where \cite{rasmussen2005gaussian}
\def\bphi{{\pmb{\phi}}}
\begin{align}
    \upredmean(\br) &= \bphi^H(\br)\Sigmaalpha\bPhi \bQ^{-1}\by, \label{eq:pred_mean}\\
    \ypredvar(\br) &= \bphi^H(\br)\Sigmaalpha\bphi(\br)-\bphi^H(\br)\Sigmaalpha\bPhi \bQ^{-1}\bPhi^H\Sigmaalpha\bphi(\br) \label{eq:pred_var},
\end{align}
where $\bphi(\br) = [\phi_1(\br),\dots,\phi_P(\br)]^T$ and
\begin{equation}
    \bQ = \left( \sigma^2 \bI + \bPhi\Sigmaalpha\bPhi^H \right).
    \label{eq:Q}  
\end{equation}
From \eqref{eq:pred_mean} and \eqref{eq:pred_var}, one may note 
that the reconstructed sound field depends on the choice of $\Sigmaalpha$, with the well-known Tikhonov estimator corresponding to the choice $\Sigmaalpha = \sigmaalpha \bI$. 
As an alternative, sound fields that consists of a sparse directivity pattern has also been modelled using a covariance structure on the form $\Sigmaalpha = \mathrm{diag}(\begin{bmatrix}\sigma_{\alpha,1}^2,\sigma_{\alpha,2}^2, \dots, \sigma_{\alpha,P}^2 \end{bmatrix})$ 
\cite{gemba2017multi}. 

\begin{figure}
    \centering
    \includegraphics[width=\linewidth]{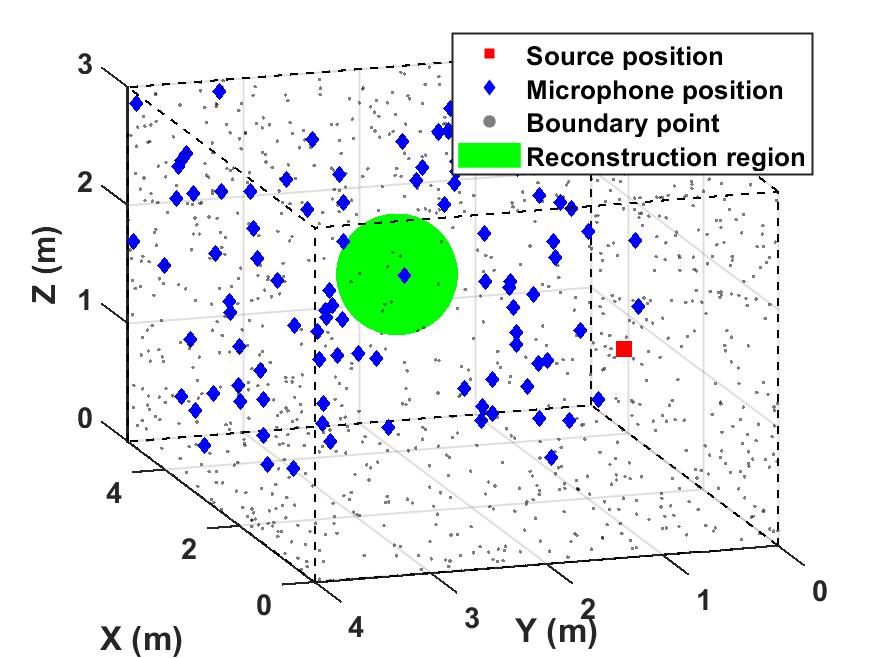}
    \caption{Illustration of the experimental setup.}
    \label{fig:setup_sim}
\end{figure}

\begin{figure}
    \centering
    \includegraphics[width=\linewidth]{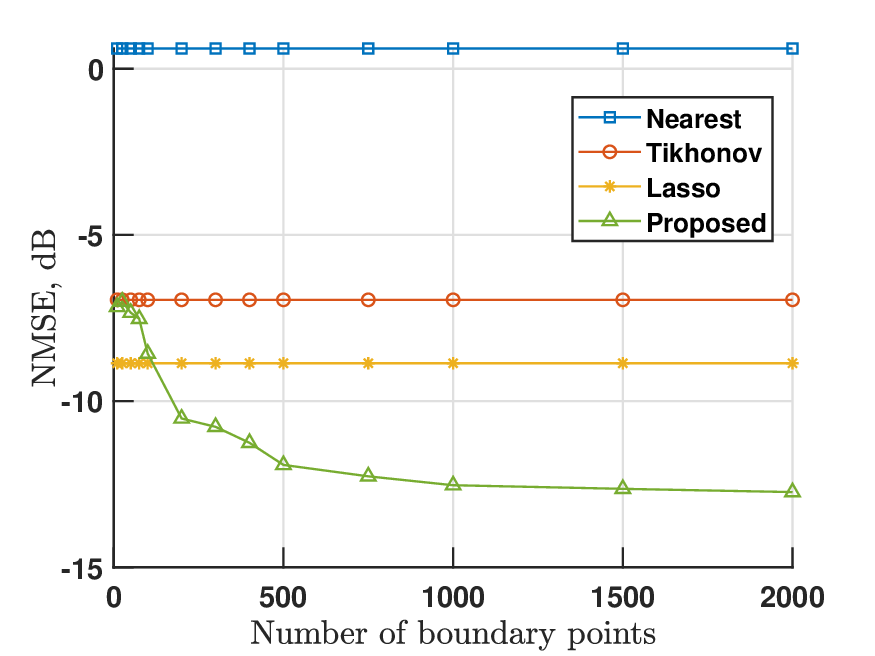}
    \caption{Illustration of the NMSE as a function of the number of boundary points used to form the regularizer.}
    \label{fig:number_boundary_points}
\end{figure}

\subsection{Introducing a boundary-informed prior distribution}
Although the reconstruction in \eqref{eq:pred_mean} inherently satisfies \eqref{eq:homogeneous_helmholtz} due to the plane wave model, we further incorporate the remaining boundary information from \eqref{eq:robin_bc} by formulating a prior distribution for \(\balpha\).
In this preliminary work, for clarity of presentation, we assume all boundary points to share the same specific impedance, i.e., $ \beta(\br) \equiv \beta$, $\forall \br \in \cB$.
By differentiating the plane-wave model along the normal, \eqref{eq:robin_bc} can be reformulated as
\begin{equation}
    \left (\beta\bPsi+\Phib\right)\balpha = 0
    \label{eq:bc_alpha}
\end{equation}
where $[\bPsi]_{b,p} = i\bk_p^T \bn(\br_b)e^{i\bk_p^T\br_b}$ and $[\Phib]_{b,p} = ike^{i\bk_p^T\br_b}$, for $p=1,\dots,P$ and $b = 1,\dots, B$, where $B$ is the cardinality of the set $\cB$. In principle, the boundary constraint \eqref{eq:bc_alpha} could be included as a linear constraint in \eqref{eq:least_squares}. However, in practice the resulting solution will only approximately satisfy the constraint in \eqref{eq:bc_alpha}, due to, for example, uncertainties in the 3D point cloud detailing the boundary. Therefore, we instead let $\left(\beta\bPsi+\Phib\right)\balpha \sim \cCN(\zerovec, I\sigmaalpha)$, which corresponds to assigning the prior 
\begin{equation}
\cCN\left(\zerovec, \sigmaalpha\left(\left(\beta\bPsi+\Phib\right)^H\left(\beta\bPsi+\Phib\right)\right)^{-1}\right) 
\end{equation}
to $\balpha$. However, when only a few boundary points are used, the matrix $(\beta\bPsi+\Phib)^H (\beta\bPsi+\Phib)$ will be rank deficient. Therefore, we propose to assign the prior distribution in \eqref{eq:alpha_prior} with the covariance defined as
\begin{equation}
    \Sigmaalpha =  \sigmaalpha\left( \bI+ \mu\left(\beta\bPsi+\Phib\right)^H\left(\beta\bPsi+\Phib\right)\right)^{-1}.
    \label{eq:prior_covariance}
\end{equation}
Note that although we in this work use a plane-wave model to represent the field, similar priors could be formulated for other basis functions using a similar approach.
Also, note that the maximum a posteriori estimate in \eqref{eq:alpha_MAP} formed using this prior coincides with the Tikhonov estimator when $\mu=0$.

\begin{figure}
    \centering
    \includegraphics[width=\linewidth]{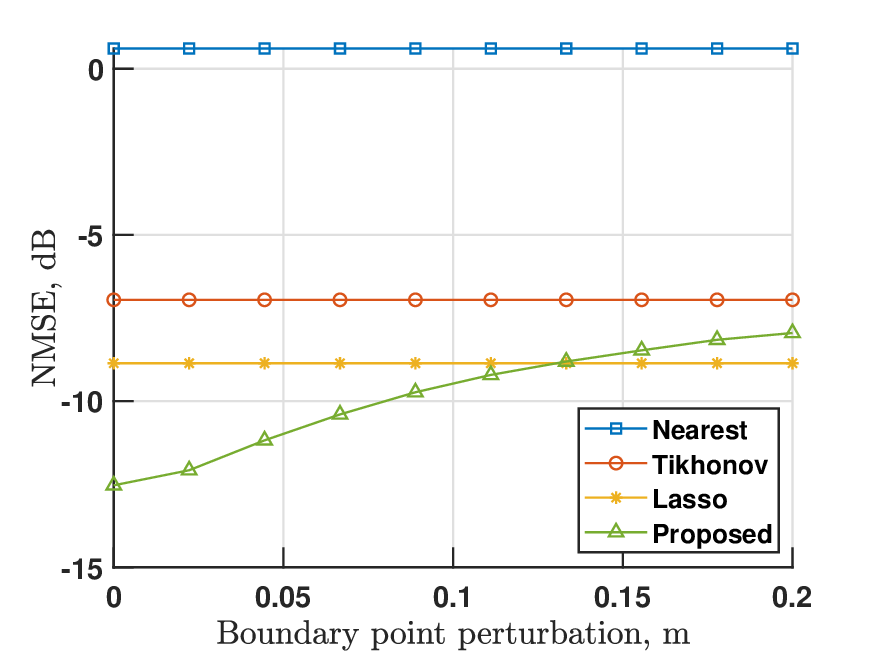}
    \caption{Illustration of the robustness with respect to constant perturbations in the assumed boundary point positions.}
    \label{fig:boundary_pos_perturb}
\end{figure}

\subsection{Joint estimation of hyperparameters and impedance boundary conditions}
In this section, we consider a Bayesian approach for estimating the unknown hyperparameters \cite{antoni2012bayesian}. 
The hyperparameters are estimated by maximizing their posterior distributions, which corresponds to maximizing the marginal likelihood (see for example \cite{rasmussen2005gaussian} for a detailed derivation)
\begin{equation}
    \argminwrt[\beta \in \CC, \sigma^2, \sigmaalpha,\mu \in \RR_+] \frac{1}{2}\by^H\bQ^{-1}\by +\frac{1}{2}\log\left(|\bQ|\right),
    \label{eq:marginal_log_likelihood_constrained}
\end{equation} 
where $|\bQ|$ denotes the determinant of $\bQ$. To account for the non-negativity constraints and introduce the fact that the parameters can span several orders of magnitude, the variables are reparametrized as $\sigma^2 = e^{a}$, $\sigmaalpha = e^b$, $\mu = e^d$, and $\beta = e^\eta$, resulting in the unconstrained problem
\begin{equation}
   J(\theta) = \argminwrt[\eta \in \CC, a,b,d \in \RR] \frac{1}{2}\by^H\bQ^{-1}\by +\frac{1}{2}\log\left(|\bQ|\right) ,
    \label{eq:marginal_log_likelihood}
\end{equation} 
where $\theta = \{a,b,d,\eta\}$.
The non-convex problem in \eqref{eq:marginal_log_likelihood} may be solved using a conjugate-gradient based solver with Polak-Ribière search directions \cite{polak1969note}, as implemented in \cite{rasmussen2010gaussian}. 
The gradients with respect to a parameter $\theta_i$ in the covariance matrix $\bQ$ can efficiently be computed by straight-forward extension of the result in \cite{rasmussen2005gaussian} for the complex-valued setting, resulting in
\begin{equation}
    \frac{\partial}{\partial \theta_i} J(\theta) = -\frac{1}{2}\mathrm{trace}\left(\left( \bxi \bxi^H-\bQ^{-1} \right) \frac{\partial\bQ}{\partial \theta_i} \right),
\end{equation}
where $\bxi = \bQ^{-1} \by$. The partial derivatives of $\bQ$ with respect to each hyperparameter are stated in the appendix.

\begin{figure}
    \centering
    \includegraphics[width=\linewidth]{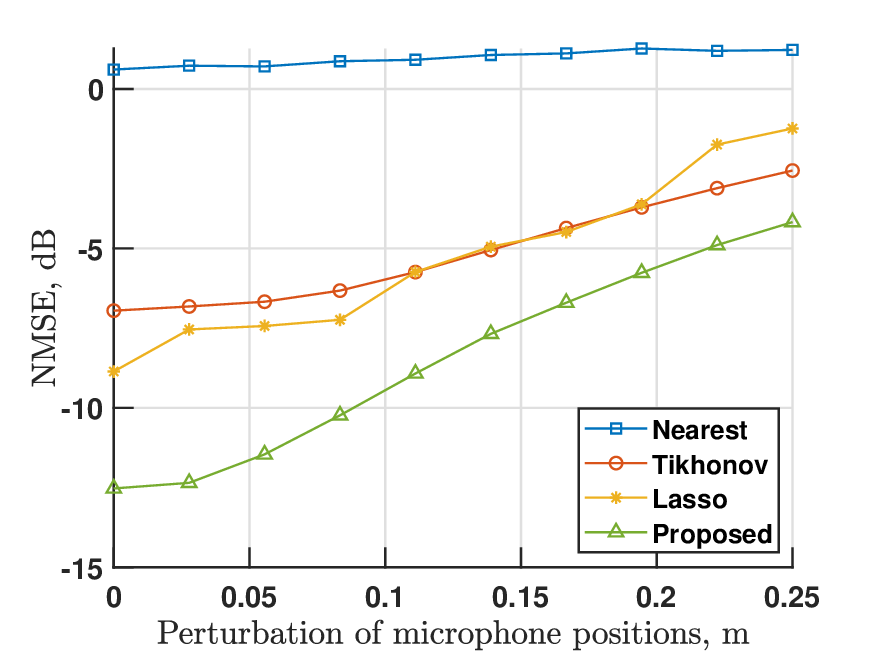}
    \caption{Illustration of the robustness with respect to constant perturbations in the assumed microphone positions.}
    \label{fig:mic_pos_perturb}
\end{figure}

\section{Numerical experiments}
\label{sect:numerical_experiments}
In the following, we want to understand the limits for when the use of a boundary-informed prior is useful and, as a result, give suggestions on the calibration accuracy that is required when collecting a real dataset for this purpose.
Therefore, sound fields are simulated in a controlled setting using the image source method (ISM) as implemented in \cite{habets2006room} with the speed of sound $343$m/s and the sampling frequency $8$kHz. 
Since the ISM simulation only is guaranteed to satisfy the boundary conditions in \eqref{eq:robin_bc} exactly as $\beta \rightarrow \infty$, that is, in the idealistic setting of lossless reflections \cite{kuttruff2016room}, the reflection coefficients are set to $0.95$ as an approximation. Note that this results in a slight mismatch between the model in Section~\ref{sect:method} and the simulation, making the evaluation setting less favorable for the proposed method.
The experimental setup is illustrated in Figure \ref{fig:setup_sim}, including $100$ microphones that are uniformly distributed in half of a shoebox room except in a spherical region of radius $0.5$ m, where instead $20$ validation positions are sampled uniformly random. Circularly symmetric Gaussian noise is added to the signals to obtain a signal-to-noise ratio of $20$dB 
for a frequency of $300$Hz.
The reconstruction is measured in terms of the normalized mean squared error, defined as
\def\uhat{\hat{u}}
\begin{align}
    \text{NMSE} = \frac{1}{JN}\sum_{j = 1}^J\sum_{n=1}^N \frac{\left|u_{j,n}-\uhat_{j,n}\right|^2}{\left|u_{j,n}\right|^2},
\end{align}
where $\uhat_{j,n}$ and $u_{j,n}$ denote the predicted and true signal for the $j$th validation point and $n$th simulation, 
for $N = 10$ Monte-Carlo simulations.
\begin{figure}
    \centering
    \includegraphics[width=\linewidth]{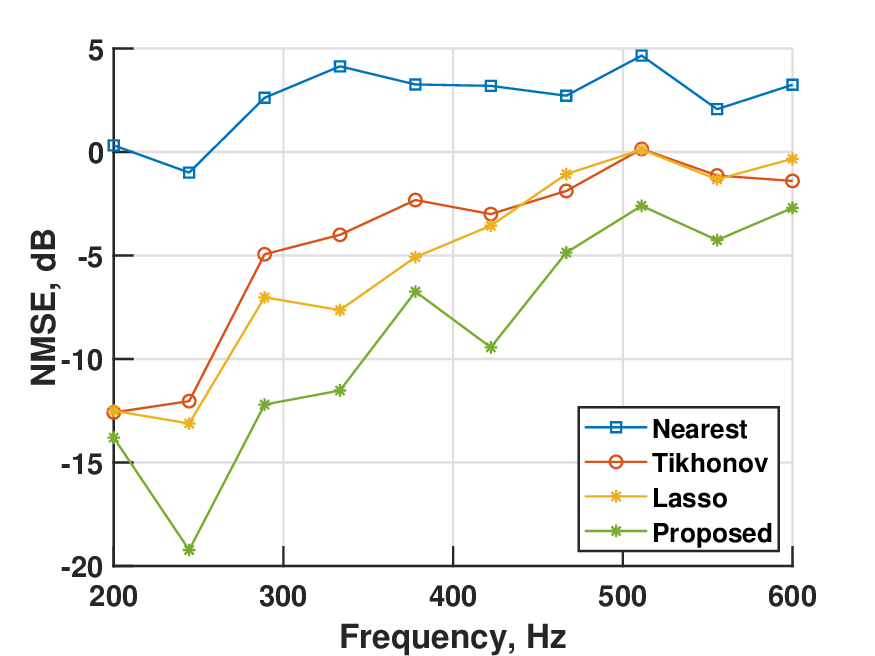}
    \caption{Illustration of the NMSE of the reconstruction for various frequencies.}
    \label{fig:frequencies}
\end{figure}
To give the results more meaning, we include three benchmark methods. First, the naive reconstruction is formed using the microphone signal located nearest to the reconstruction position, denoted Nearest. Secondly, the Tikhonov estimator is constructed by setting $\mu=0$ in \eqref{eq:prior_covariance}, which, compared to the proposed method provides an interpretation of the impact of the boundary information in the prior covariance in \eqref{eq:prior_covariance}. Finally, also the Lasso estimate is obtained by adding the regularizing term $\lambda||\balpha||_1$ to \eqref{eq:least_squares}. For all methods, 1000 plane waves distributed on a Fibonacci lattice are used \cite{gonzalez2010measurement}.
Firstly, we study how the number of boundary points affects the reconstruction. The boundary points are uniformly sampled on the surface of the shoebox room with outward normals, as illustrated in Figure \ref{fig:setup_sim}, with the results shown in Figure~\ref{fig:number_boundary_points}. Notable is that the reconstruction error decreases rapidly when including the first few hundreds of points and flats out when about 1000 boundary points are used, which therefore is used in the following experiments. Of practical interest is also studying the robustness with respect to the assumption of the boundary point and microphone positions. In order to do so, a constant perturbation of random direction is introduced to each boundary point and microphone position in Figures~\ref{fig:boundary_pos_perturb} and \ref{fig:mic_pos_perturb}. In Figure~\ref{fig:boundary_pos_perturb}, it is worth noting that Proposed obtains a lower reconstruction error as compared the other methods for boundary position errors up to as high as $1$~dm. This result open up the possibility of being able to use a point cloud representation of the boundary obtained from everyday sensors such as, for example, the camera sensor in a smartphone, a case 
which will be examined in more detail in further work. 
Figure~\ref{fig:mic_pos_perturb} illustrates that Proposed achieves a lower NMSE than the other methods even in the presence of large errors in the assumption of microphone position. This is not 
obvious,
since Proposed includes an assumption of the microphone position in both the observation model and the prior, while the other models only include the assumption of the microphone positions in the observation model. However, the result indicate that the error due to the error in the observation model dominates the error of the reconstructed sound field in this case. 
In general, the results indicate that the boundary information is of significance value, even when only partial or uncertain information of the boundary is available, which is also confirmed for a wide frequency range in Figure~\ref{fig:frequencies}.

\section{Conclusion}
We introduce a boundary-informed prior distribution for the reconstruction of a sound field from microphone measurements. Compared to simulation-based approaches, which require the full geometry, we show that boundary information is useful also in settings where the point cloud representation is coarse or uncertain. Simulated numerical experiments in reverberant rooms both illustrate the great potential benefit of this information, and serves as a guideline for the required accuracy when collecting data for a real experiments.

\appendix
\section*{Gradients of $\bQ$ with respect to hyperparameters}
\def\bB{{\mathbf B}}
The gradient of $\bQ$ in \eqref{eq:Q} with respect to $a$ and $b$ are given by $\frac{\partial}{\partial a} \bQ = \sigma^2I $ and $\frac{\partial}{\partial b} \bQ  = \Phi\Sigmaalpha \Phi^H$, respectively. 
Furthermore, the gradients with respect to $d$ is given by $\frac{\partial}{\partial d}  \bQ = -\sigmaalpha \mu \bB^H \bB$, where
\begin{align*}
    \bB &=  \left(\beta\bPsi + \Phib\right)\left(I+ \mu \left(\beta\bPsi + \Phib\right)^H\left(\beta\bPsi + \Phib\right)\right)^{-1}\bPhi^H.
\end{align*}
To compute the gradient with respect to the complex-valued variable $\eta$, we note that $\bQ$ is not a holomorphic function with respect to $\eta$ due to the conjugation of $\eta$ in \eqref{eq:prior_covariance}. Therefore, the derivative does not exist in the classical sense, 
but may, 
the real and imaginary part, which, 
using Wirtinger calculus, be expressed as the derivative with respect to $\eta^*$ 
(see, e.g., 
\cite{haykin2002adaptive}), 
yielding
\begin{align}
    2\frac{\partial}{\partial \eta^*} \bQ &= -2\sigmaalpha\mu\bC^H \left(|\beta|^2\bPsi^H\bPsi+\beta^*\bPsi^H\Phib\right) \bC
\end{align}
where 
 $   \bC = \left( I+ \mu (\beta\bPsi+\Phib)^H(\beta\bPsi+\Phib)\right)^{-1} \bPhi^H$.

\bibliographystyle{IEEEbib}
\bibliography{export}

\end{document}